\documentclass[12pt]{article}

\usepackage{arxiv}
\usepackage{booktabs}
\usepackage{graphicx}
\usepackage{url}

\AtBeginDocument{%
  }

\begin{document}

\title{Understanding Housing and Homelessness System Access by Linking Administrative Data}


\author{
Geoffrey Messier\\
Electrical and Software Engineering\\
University of Calgary\\
\texttt{gmessier@ucalgary.ca}
\and
Sam Elliott\\
Calgary Homeless Foundation\\
\texttt{same@calgaryhomeless.com}
\and
Dallas Seitz\\
Psychiatry\\
University of Calgary\\
\texttt{dallas.seitz@ucalgary.ca}
}

\date{}

\maketitle


\begin{abstract}
This paper uses privacy preserving methods to link over 235,000 records in the housing and homelessness system of care (HHSC) of a major North American city.  Several machine learning pairwise linkage and two clustering algorithms are evaluated for merging the profiles for latent individuals in the data.  Importantly, these methods are evaluated using both traditional machine learning metrics and HHSC system use metrics generated using the linked data.  The results demonstrate that privacy preserving linkage methods are an effective and practical method for understanding how a single person interacts with multiple agencies across an HHSC.  They also show that performance differences between linkage techniques are amplified when evaluated using HHSC domain specific metrics like number of emergency homeless shelter stays, length of time interacting with an HHSC and number of emergency shelters visited per person.  
\end{abstract}







\section{Introduction}
\label{sec:Intro}

Homelessness is one of the most significant challenges facing society today.   In Canada, more than 25,000 people in 61 communities experienced homelessness on a single day in 2018 and an estimated 235,000 people in Canada experience some form of homelessness each year \cite{dionne2023}.  It is also now well recognized that experiencing homelessness is the result of a person facing a variety of complex mental health, physical health, income, trauma and addiction related challenges \cite{czechowski2022}.  The system caring for people experiencing homelessness is similarly complex.  The housing and homelessness system of care (HHSC) in Calgary, Canada in 2024 had over 24 agencies serving people experiencing homelessness \cite{calgaryhomelessfoundation2024}.

The large number of agencies and services working in the homelessness sector has been identified as one of the primary challenges in helping people exit homelessness in an effective and coordinated way \cite{hambrickjr.2000}.  It is not uncommon for people experiencing homelessness to interact with multiple agencies \cite{jadidzadeh2020}.  Therefore, understanding a person's experience of homelessness depends on understanding their experience with the multiple agencies that make up the overall system of care.

Most HHSC agencies record administrative data when people access their services.  However, a major challenge is the fragmentation of that data.  HHSC agencies often use incompatible information technology (IT) systems to store their administrative data and people experiencing homelessness are not provided with HHSC-wide unique identifiers.  Before administrative data can be used to understand a person's interaction with an entire HHSC, the administrative data records within that HHSC must be linked and the absence of a unique identifier means that this linkage must use identifying information like names and birth-dates.    This linkage must be accurately performed in spite of errors in how identifying information is recorded from one agency to the next.  Linkage must also respect HHSC user privacy by not exporting plain text identifying information beyond agency boundaries.

Fortunately, there is a rich body of work on record linkage and de-duplication using error prone identifying information that can be brought to bear on this problem \cite{binette2022}.  A subset of these methods are designed to preserve privacy during the linkage operation \cite{vatsalan2013}.  The nature of HHSC access has a direct impact on which linkage method is the most appropriate.  This paper will evaluate the linked data set not just with traditional machine learning metrics that focus on linkage algorithm performance.  The linked data sets produced by different linkage methods will also be compared using HHSC domain specific metrics like number of emergency shelter stays, length of time interacting with the HHSC and number of shelters visited by a person.  These {\em domain specific} metrics will demonstrate how performance differences between algorithms manifest in the metrics relevant to HHSC operators and program administrators.  Domain specific metrics will be provided for the entire cohort of HHSC users and the top 5th percentile of heaviest system users since it is well established that a small proportion of HHSC users account for the highest system use \cite{kuhn1998,culhane2007,aubry2013,kneebone2015}.  

The primary contribution of this paper is to demonstrate that privacy preserving record linkage methods are a practical way to understand HHSC wide system interaction and can change our perspective on how much people use HHSC services. A second contribution is to demonstrate the importance of considering both machine learning metrics and domain specific metrics when evaluating the performance of privacy preserving linkage algorithms.  Results will be presented demonstrating that the relative performance differences between algorithms are amplified when domain specific HHSC metrics are used to compare those algorithms.

\section{Related Work}
\label{sec:RelatedWork}

The excellent survey article by Binette and Steorts succinctly summarizes the different options available when linking records \cite{binette2022}.  Many of these methods are unsupervised and useful when no training data set is available with labels indicating when a match has occurred.  However, as will be discussed in Section~\ref{sec:Data}, the data set used for this project includes a set of manually linked profiles.  This allows us to adopt a supervised machine learning approach for identifying pairwise matches between profiles followed by clustering as a post-processing step to gather pairs of profile matches into groups that represent a single latent individual.  

The two step entity resolution approach of first detecting pairwise matches and then clustering those matches together is summarized by Christophides, et. al. \cite{christophides2020}.  The supervised machine learning approach we adopt for detecting pairwise matches most closely resembles the work in \cite{cochinwala2001,ventura2015}.  While considerable progress has been made applying deep learning techniques to entity matching \cite{mudgal2018}, the methodology in this paper will be restricted to classical machine learning models since we demonstrate that these are more than adequate to achieve excellent performance for our application.  Regarding clustering, as noted in \cite{christophides2020, hassanzadeh2009a}, simply applying the transitive closure on a set of identified pairwise matches will maximize recall but is extremely sensitive to false positive matching errors.  As a result, we explore the more noise robust clustering approaches \cite{hassanzadeh2009a} as described in Section~\ref{ssec:Cluster}.

A priority for this project is protecting the privacy of the people in the housing and homelessness data sets being linked.  This requires special consideration since linkage will be performed using names and birthdates.  Following the taxonomy of \cite{vatsalan2013}, we adopt the approximate matching privacy preserving record linkage (PPRL) approach where each partner or agency providing data first scrambles people's identifying information using the Bloom filter approach \cite{schnell2009}.  This preserves privacy while still allowing the similarity of two records to be evaluated by calculating the Dice coefficient for the records' corresponding Bloom filtered identifying information.  Pairwise matches are detected using machine learning as in \cite{cochinwala2001} except that identifying information similarity is calculated using Dice coefficients rather than a direct comparison of the plain text identifying fields.

As indicated in Section~\ref{sec:Intro}, this study will go further than traditional machine learning performance metrics to evaluate the effect of linkage errors using metrics that are relevant for the application space of the data (in this case, housing and homelessness).  In this same spirit, some notable studies do investigate the impact of linkage errors and different linkage methodologies using domain specific metrics.  This includes \cite{prindle2023} for California demographics information, \cite{tancredi2020} for fatalities in the Syrian conflict, \cite{gutman2016} on the length of time to connect HIV+ inmates to care, \cite{ventura2015} on patent inventor data statistics and \cite{gutman2013} on end of life medical costs.  However, none of these studies work with housing and homelessness system utilization data.  Other studies explore application specific metrics post-linkage but do not investigate the impact of linkage error on those metrics.  This includes a study linking health and homelessness data in Illinois \cite{trick2021} and records from multiple health sites in Chicago \cite{kho2015}.   Finally, many linkage studies limit their discussion to confusion matrix based machine learning performance metrics only.  This includes an application of linkage methods to healthcare in Malawi \cite{dixon2023}, medical and emergency department records \cite{redfield2020}, Canadian diagnostic imaging respositories \cite{nagels2019}, Brazilian beef cattle farm data \cite{aiken2019}, traumatic injurty brain registries \cite{kesinger2017}, El Salvador homicide data \cite{sadinle2014}.

\section{Data Set}
\label{sec:Data}

The data set utilized for this study consists of housing and emergency homeless shelter access records for the HHSC in Calgary, Canada.  Records for 20 different emergency shelters and over 2,000 housing programs are included for the date range October 30, 1995 to December 24, 2023.  The data is provided by the Calgary Homeless Foundation, the system coordination organization for the Calgary HHSC which collects data from all Calgary HHSC programs and agencies that receive provincial government funding support.  The anonymization and security protocols used to handle the data have been approved by the {\tt $<$Anonymized for Review$>$} ethics board.

The first component of the data set is a list of first names, last names, birth months, birth days and birth years of each person accessing each of the programs in the data set.  To preserve privacy, this identifying information is scrambled by the agency using Bloom filtering \cite{schnell2009} prior to release to the researchers.  As demonstrated in \cite{schnell2009}, the parameters of the Bloom filter scrambling can have an impact on linkage performance.  To explore the trade offs of increasing Bloom filter vector length, the identifying information was Bloom filter scrambled using 32 and 64 bit vectors, both generated using 2-ary Q-grams. The scrambled identifying information for each person is also indexed by a non-identifying ID number.  

The second component of the data set contains anonymized HHSC service access records.  Each record consists of the non-identifying ID number of the person accessing a shelter or housing program and the date of the shelter stay or program access.  

Each entry in the Bloom filter scrambled list of names and birth-dates is referred to as a {\em profile}.  The first time a person accesses a new program or shelter within the HHSC, a new profile is created for them in the CHF data.  Since the CHF does not automatically merge the data it collects from different agencies and programs, a new profile is created for the same person each time that person accesses a new program in the Calgary HHSC.   Therefore, the number of profiles for a person is equal to at least the number of different agencies they have interacted with.  In some cases, the same person may have multiple profiles at the same agency if spelling or data entry errors are made when the person accesses services.  There are 235,230 profiles in the data set.  

As described in more detail in Section~\ref{sec:Methods}, a pairwise comparison between profiles will be conducted using a machine learning model in order to identify {\em clusters} of profiles that correspond to the same person.  This person is sometimes referred to as the {\em latent person} or {\em latent entity} in record linkage literature.  In order to train the supervised machine learning model that identifies links between profiles, the CHF also provided an anonymized table of profiles that were manually linked by CHF staff.  

To create this manually linked table, a profile is randomly selected from the full CHF profile list and the top 10 best matches for that profile are presented to the staff member for inspection.  This list of best candidate matches is created by  concatenating first name, last name and date of birth information into a single string for all profiles.  The edit distance is then calculated between the concatenated string of the randomly selected profile and the concatenated strings of all the other profiles in the data set.  The Levenshtein or {\em edit distance} between two strings equals the minimum number of character edits required to transform one string into the other \cite{elmagarmid2007}.  The top profiles displayed for inspection by the staff person are the ones that are the smallest edit distance from the randomly selected profile.  

A randomly selected profile and any profiles manually matched with it form a single {\em ground truth cluster}.    The smallest possible ground truth cluster is a single profile where no matches to other profiles were identified by the CHF staff.  A total of 326 profiles were randomly selected from the full profile list and manually linked to 775 other profiles from that list.  This means the manual link table groups a total of 1,101 profiles into 326 ground truth clusters.  The edit distances for between each identifying information field for each manual match are also recorded and provided to the researchers.

\section{Synthetic Data Generation}
\label{sec:Synth}

In order to increase the amount of data available for machine learning linkage model training, a synthetic data set is generated using 4,750 unique records with non-empty entries from the public domain Freely Extensible Biomedical Record Linkage 4 (FEBRL 4) data set \cite{christen2008}.  Each of the 4,750 unique synthetic profiles from the  FEBRL 4 list will be referred to as {\em original synthetic profiles}.  This section will describe how duplicates will be generated for these original synthetic profiles to create synthetic clusters with similar statistics to the manually matched ground truth clusters described in Section~\ref{sec:Data}.

The cluster size distribution of the manually matched table is empirically determined and is shown in Table~\ref{tb.ClstrSz}.  For each synthetic original profile, a random number of duplicates is generated from this empirical manual cluster size distribution.  The result is a total of 16,058 synthetic profiles that are grouped into 4,750 clusters  with size distributions also shown in Table~\ref{tb.ClstrSz}.

\begin{table}[htbp]
\centering
  \begin{tabular}{ccc}
    \toprule
     & Manual Match  & Synthetic \\
    Cluster Size & No.~Clusters & No.~Clusters\\
    \midrule
1 & 60 / 326 (18.40\%) & 848 / 4,750 (17.85\%) \\
2 & 96 / 326 (29.45\%) & 1435 / 4,750 (30.21\%) \\
3 & 53 / 326 (16.26\%) & 750 / 4,750 (15.79\%) \\
4 & 34 / 326 (10.43\%) & 499 / 4,750 (10.51\%) \\
5 & 29 / 326 (8.90\%) & 447 / 4,750 (9.41\%) \\
$>$5 & 54 / 326 (16.56\%) & 771 / 4,750 (16.23\%) \\
\bottomrule
\end{tabular}
\caption{Manually matched and synthetic cluster sizes.}
\label{tb.ClstrSz}
\end{table}

As noted in Section~\ref{sec:Intro}, one of the challenges of linking profiles using identifying information is that there will be errors in how identifying information is recorded in different places for the same person.  These error statistics for the manually matched profiles can be illustrated by the edit distance and Dice coefficient values between manually linked profiles.  The top 10 most common error patterns observed in the manually matched data are shown in Table~\ref{tb.ErrMan} where the Dice coefficients are calculated using the 64~bit Bloom filter vectors.  The most common error pattern (53.81\% of profiles) is an identical match where the identifying information was recorded correctly for two duplicate profiles.  Each non-identical error pattern is approximately equally likely and predominately shows error in a single identifying field.  Anecdotally, CHF staff observed during the manual linkage process that edit distance errors of 3 or less are most often typos (ie. "Geoff" vs "Jeoff") and or short form name variations (ie. "Geoffrey" vs "Geoff").  Longer errors appear to be intentional variations provided by people when accessing different services.

\begin{table}[htbp]
\centering
  \begin{tabular}{c|ccccc}
    \toprule
    No.~Rec. & First Name & Last Name & DOB Day & DOB Month & DOB Year \\
    \midrule
417 / 775 (53.81\%) & (0,1.00) & (0,1.00) & (0,1.00) & (0,1.00) & (0,1.00) \\
30 / 775 (3.87\%) & (0,1.00) & (1,0.85) & (0,1.00) & (0,1.00) & (0,1.00) \\
29 / 775 (3.74\%) & (3,0.58) & (0,1.00) & (0,1.00) & (0,1.00) & (0,1.00) \\
21 / 775 (2.71\%) & (1,0.82) & (0,1.00) & (0,1.00) & (0,1.00) & (0,1.00) \\
19 / 775 (2.45\%) & (0,1.00) & (0,1.00) & (1,0.39) & (0,1.00) & (0,1.00) \\
17 / 775 (2.19\%) & (0,1.00) & (5,0.52) & (0,1.00) & (0,1.00) & (0,1.00) \\
16 / 775 (2.06\%) & (0,1.00) & (2,0.65) & (0,1.00) & (0,1.00) & (0,1.00) \\
16 / 775 (2.06\%) & (4,0.48) & (0,1.00) & (0,1.00) & (0,1.00) & (0,1.00) \\
12 / 775 (1.55\%) & (0,1.00) & (0,1.00) & (0,1.00) & (0,1.00) & (1,0.67) \\
12 / 775 (1.55\%) & (6,0.58) & (0,1.00) & (0,1.00) & (0,1.00) & (0,1.00) \\
11 / 729 (1.51\%) & (2,0.605) & (0,1.000) & (0,1.000) & (0,1.000) & (0,1.000) \\
\bottomrule
\end{tabular}
\caption{Top 10 most common manual match pairwise (Edit Distance, Dice Coefficient) error patterns.}
\label{tb.ErrMan}
\end{table}

When generating a synthetic duplicate profile, an edit error pattern is drawn from the empirical manually linked error distribution shown in Table~\ref{tb.ErrMan} and the synthetic duplicate is distorted appropriately.  If an error is being added to a synthetic duplicate name,  the required number random characters to achieve the edit distance are added to, deleted from or substituted within the original name.  The choice to add, delete or substitute characters is made randomly and with equal likelihood.  The required number of date of birth digits to achieve a certain edit distance are altered while still ensuring the altered number is a valid birth year, day or month as required.  The top 10 most common error patterns for these synthetic duplicates are shown in Table~\ref{tb.ErrSyn}.

There is good agreement between the edit distance and Dice coefficient error patterns for the manually matched and synthetic duplicate records shown in Tables~\ref{tb.ErrMan} and \ref{tb.ErrSyn}, respectively.  The order of some of the less common error patterns do not match exactly due to statistical variation in the number of duplicates generated for the synthetic table.  When the edit distance values for the synthetic and manual tables match, the Dice coefficient values are close but not exactly the same.  This small variation is because the Dice coefficient values are not just a function the number of characters different between strings but also the length of those strings.  Therefore, the variation is because the name lengths in the synthetic and manual original records are not always the same.

\begin{table}[htbp]
\centering
  \begin{tabular}{c|ccccc}
    \toprule
    No.~Rec. & First Name & Last Name & DOB Day & DOB Month & DOB Year \\
    \midrule
6,039 / 11,308 (53.40\%) & (0,1.00) & (0,1.00) & (0,1.00) & (0,1.00) & (0,1.00) \\
456 / 11,308 (4.03\%) & (3,0.57) & (0,1.00) & (0,1.00) & (0,1.00) & (0,1.00) \\
435 / 11,308 (3.85\%) & (0,1.00) & (1,0.80) & (0,1.00) & (0,1.00) & (0,1.00) \\
331 / 11,308 (2.93\%) & (1,0.79) & (0,1.00) & (0,1.00) & (0,1.00) & (0,1.00) \\
315 / 11,308 (2.79\%) & (0,1.00) & (0,1.00) & (1,0.33) & (0,1.00) & (0,1.00) \\
247 / 11,308 (2.18\%) & (0,1.00) & (5,0.55) & (0,1.00) & (0,1.00) & (0,1.00) \\
242 / 11,308 (2.14\%) & (0,1.00) & (2,0.69) & (0,1.00) & (0,1.00) & (0,1.00) \\
235 / 11,308 (2.08\%) & (4,0.54) & (0,1.00) & (0,1.00) & (0,1.00) & (0,1.00) \\
183 / 11,308 (1.62\%) & (6,0.57) & (0,1.00) & (0,1.00) & (0,1.00) & (0,1.00) \\
175 / 11,308 (1.55\%) & (2,0.66) & (0,1.00) & (0,1.00) & (0,1.00) & (0,1.00) \\
\bottomrule
\end{tabular}
\caption{Top 10 most common synthetic pairwise (Edit Distance, Dice Coefficient) error patterns.}
\label{tb.ErrSyn}
\end{table}

\section{Linkage and Clustering Methodology}
\label{sec:Methods}

Following the two step entity resolution process described in \cite{christophides2020}, profiles will be associated with latent people by first conducting a pairwise comparison of all profiles in the data set.   When two profiles are compared, the similarity of their Bloom filter encoded identifying information fields will be calculated using the Dice coefficient \cite{schnell2009}. A machine learning model utilizes the Dice coefficient value(s) to determine if the profiles belong to the same person.  Once the pairwise linkages have been determined, a clustering algorithm is used to gather them into groups that represent the same latent person.  Details of the Dice coefficient calculation are provided in Section~\ref{ssec:Dice} and the creation of the training and testing data sets is discussed in Section~\ref{ssec:PairMLTT}. The pairwise match machine learning models chosen for the study are discussed in Section~\ref{ssec:PairMLModels} and the clustering algorithms are described in Section~\ref{ssec:Cluster}.

\subsection{Dice Coefficient Calculation}
\label{ssec:Dice}

For the machine learning models in Section~\ref{ssec:PairMLModels} able to work with multiple input data features, the Dice coefficient is calculated separately for the first name, last name, birth day, birth month and birth year fields.  These five coefficient values are the machine learning model inputs.  Section~\ref{ssec:PairMLModels} also presents a simple threshold method that requires a single Dice coefficient value that quantifies the overall similarity of the name and date of birth fields.  Let profile $i$ consist of the $L=5$ Bloom filter encoded fields $\{f_{i,0}, f_{i,1}, \ldots f_{i,L-1}\}$ where $i=0,1$ represents the two profiles being compared.  The overall similarity of the profiles is quantified by

\begin{equation}
D_{\rm All} = \frac
{2\left( \sum_{l=0}^{L-1} h(f_{0,l},f_{1,l}) \right)}
{ \sum_{l=0}^{L-1}\left(|f_{0,l}| + |f_{1,l}|\right) }
\label{eq.Dall}
\end{equation}

\noindent
where $h(a,b)$ equals the number of bit positions equal to 1 in binary vector $a$ and $b$ and $|a|$ is the length of binary vector $a$.

\subsection{Pairwise Match Training and Test}
\label{ssec:PairMLTT}

The machine learning models used to declare a pairwise match between two profiles are trained, hyper-parameter tuned and tested first using the synthetic data set described in Section~\ref{sec:Synth}.  With $K_S$ = 16,058 synthetic records, there are $N_S=K_S(K_S-1)/2$ = 128,921,653 pairwise comparison examples.  Of those comparisons, 31,769 correspond to positive matches giving a data set imbalance of 0.0247\% (1:4,053).  This extreme imbalance is typical of pairwise record linkage applications where every record in a data set is compared with every other record.  The $K_M$ = 1,101 manually linked records were used also used for model testing and corresponds to a data set of $N_M$ = 605,550 comparisons with an imbalance of 0.3617\% (1:276).

Randomized stratified splitting is used to divide the synthetic pairwise data set into training and testing sets with a 70\%/30\% ratio.  Hyper-parameter combinations are evaluated using stratified K-fold cross validation on the training set with 5 folds \cite{ojala2009}.  Once hyper-parameter tuning is complete, the model is trained on the full training set and separately tested on the synthetic test set and the entire manually linked pairwise data set.  The synthetic and manually linked data set test performance results are reported separately.  

Models with the best test performance on the synthetic and manual data sets are selected to link the full $K$ = 235,230 profile data set described in Section~\ref{sec:Data}.  The approximately 27.7 billion pairwise comparisons were conducted on a 32 CPU core, 128~GB, 4.5~GHz personal computer using the dask parallelization library \cite{daskdevelopmentteam2016}.

\subsection{Pairwise Match Models}
\label{ssec:PairMLModels}

For this study, the simple threshold, logistic regression, decision tree and multi-layer perceptron neural network machine learning models were evaluated for detecting pairwise profile matches.  With the exception of the threshold model, all models were implemented using the scikit-learn (sklearn) python library, version 1.4.1 \cite{scikit-learn}.  

Once these algorithms are used to identify pairwise matches, the clustering algorithms described in more detail in Section~\ref{ssec:Cluster} are used to create groups of pairwise matches that are associated with latent individual people in the data set.  A key input for these clustering algorithms is a soft metric that represents confidence in each pairwise match.  These confidence values are determined using the {\tt predict\_proba()} library function for the sklearn models.  For the threshold model, linkage confidence is equal to the Dice coefficient value $D_{\rm All}$ for positive matches and $1-D_{\rm All}$ for negative matches.

The hyperparameter tuning and implementation details for each pairwise linkage model are as follows.

\paragraph{Threshold}  The threshold pairwise match model calculates (\ref{eq.Dall}) for two profiles and declares a match if the Dice coefficient value is greater than or equal to a threshold $\beta$.  Hyperparameter tuning of this model involves investigating different values of $\beta = (0,1]$ to achieve different trade-offs between precision and recall.  

\paragraph{Logistic Regression (LR)} LR hyperparameter tuning was conducted for class weights 0.01 to 500, L1 and L2 regularization penalties, regularization strength $C$ values 1e-5 to 1, maximum iteration thresholds 50 to 200 and solvers {\tt liblinear}, {\tt lbfgs} and {\tt sag}.  

\paragraph{Decision Tree}  The sklearn implementation of the decision tree algorithm is utilized without scaling.  Hyper parameter tuning conducted for minimal cost-complexity pruning parameters of 1e-4, 1e-5 and for maximum leaf nodes of 5 and 6.  These parameters are selected to explore decision trees that provide a compromise between performance and interpretability.  

\paragraph{Multi-layer Perceptron (MLP)} MLP hyperparameter tuning is conducted for the {\tt lbfgs}, {\tt sgd} and {\tt adam} solvers.  The alpha L2 regularizaton parameter is swept over the range 1e-7 to 1 in multiples of 10 and hidden layer tuples (15,), (10,), (20,), (10,3) and (6,2) are considered.  These are modestly sized models but were able to achieve strong performance, as demonstrated in Section~\ref{sec:Results}.

\subsection{Clustering Algorithms}
\label{ssec:Cluster}

Once pairwise matches have been declared, clustering algorithms are necessary to gather them into groups associated with latent people in the data set.  In general, the pairwise profile links can be represented as a similarity graph where profiles are nodes and a link between two profiles is represented as a weighted edge \cite{hassanzadeh2009a}.  A clustering algorithm identifies groups of nodes on the similarity graph that represent the same person.  For this study, the CENTER and MERGE-CENTER \cite{hassanzadeh2009a} algorithms are utilized.

It is not possible to evaluate the overall performance of a clustering algorithm on the full $K$ = 235,230 profile data set since manually verifying each of the 27.7 billion comparisons is not feasible.  However, clustering algorithm performance can be estimated by evaluating the portion of clustering results that overlap with the subset of $K_M$ = 1,101 profiles that have been manually matched.  Specifically, the definitions of cluster precision and recall are adapted from \cite{hassanzadeh2009a} to assess the performance of the full data set clustering exercise using the 1,101 manually identified clusters.  Each ground truth cluster from the manually matched data set, $g$, is associated with the estimated cluster, $f(g)$, from the full data set that shares the the highest number of common profiles.  Precision and recall then equal $|f(g)\hat g|/|f(g)|$ and $|f(g)\hat g|/|g|$, respectively.

\section{Housing/Homelessness System of Care Utilization Analysis}

As discussed in Section~\ref{sec:RelatedWork}, the selection of a record linkage scheme should not be based solely on confusion matrix based machine learning parameters such as precision, recall or accuracy.  Linkage algorithms should also be compared and evaluated using performance metrics relevant to the domain of the data itself.   To this end, the results in Section~\ref{sec:Results} will include metrics relevant to people who work and design programs within an HHSC.

Similar to many jurisdictions, the Calgary HHSC offers a range of supportive housing programs and models that go beyond traditional emergency homeless shelter services.  However, our analysis will be restricted to emergency shelter data since shelters are still the agencies within an HHSC providing services with the lowest barrier to access.  We will determine a person's total number of shelter stays and the total number of days between a their first and last day of interaction with shelter services (referred to as that person's {\em tenure}).  When combined together, these two metrics give an impression of whether a person is a long term, continuous HHSC user (sometimes referred to as a {\em chronic} user \cite{culhane1994}) or a long term, infrequent user (sometimes referred to as an {\em episodic} user \cite{culhane1994}).  We will also determine the number of shelters a person accesses during their tenure. This gives an idea of how interconnected HHSC services are and was a metric of particular interest during the COVID-19 pandemic \cite{jadidzadeh2020,messier2024b}.

\begin{figure}[htbp]
\centering
\includegraphics[width=5in]{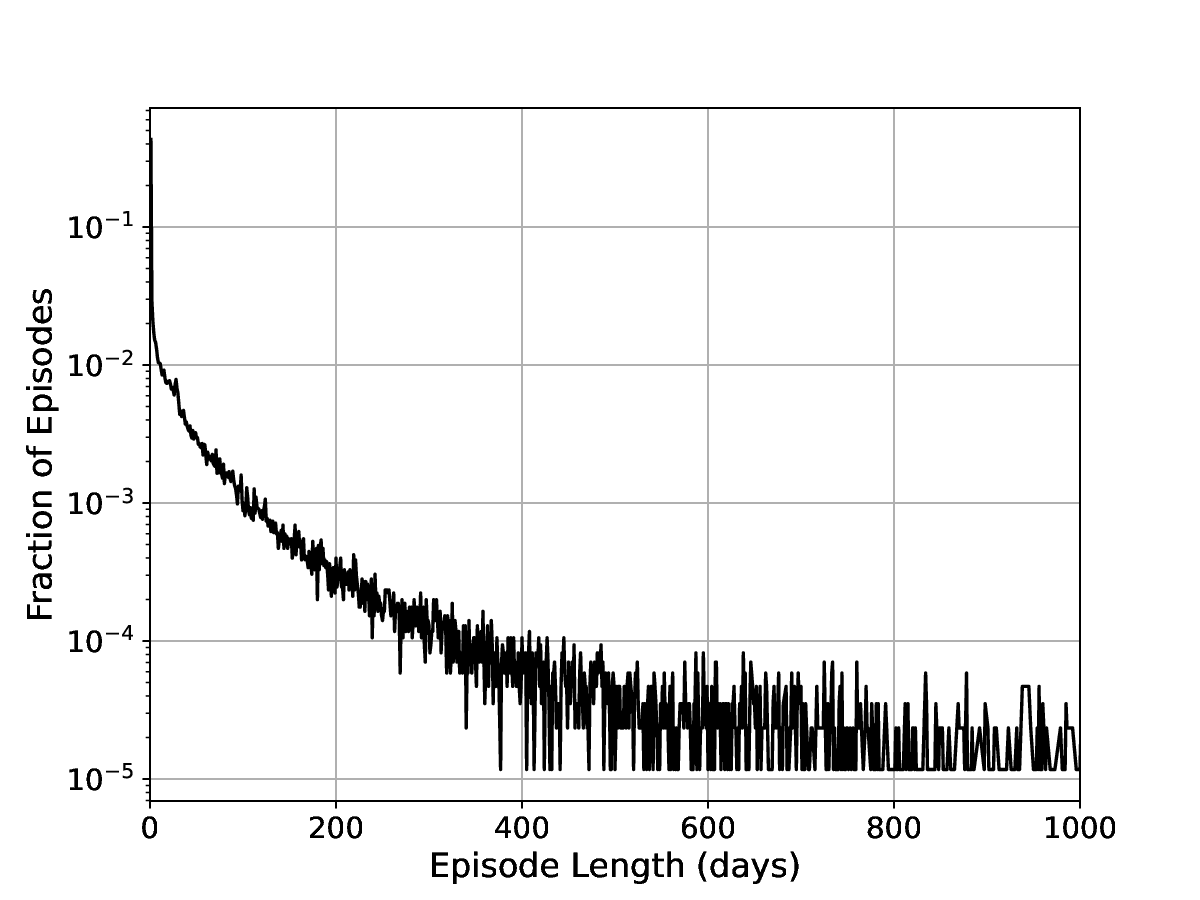}
\caption{Shelter access episode length distribution.}
\label{fg.EpiDist}
\end{figure}

These metrics will be calculated using an anonymized data set of emergency shelter access records for 12 shelters from January 1, 2016 to December 31, 2021.  This data set contains 85,229 individual shelter access records where each entry is an episode of shelter use for a particular user profile.    An episode is defined as a period of consecutive shelter stays with gaps between stays of less than 30 days.  The distribution of the length of shelter episodes are shown in Figure~\ref{fg.EpiDist}.  Note that this figure is determined using episodes from individual profiles that have not yet been linked into clusters associated with latent people in the data.

\section{Results}
\label{sec:Results}

\subsection{Linkage and Clustering}
\label{ssec:AlgPerf}

\newcommand{\Thresh}[1]{Thresh$_{#1}$}

The pairwise linkage performance results for the testing described in Section~\ref{ssec:PairMLTT} are shown in Table~\ref{tb.PairPerf} for the 32 and 64 bit length Bloom filter vectors.  The table reflects the best performance achieved during the hyperparameter tuning with the final hyperparameters indicated.  While most models with both vector lengths give acceptable performance, slightly better performance is achieved for the longer 64 bit vectors which is consistent with \cite{schnell2009}.  Three models are chosen to link the full profile data set using the 64 bit Bloom filter vectors: \Thresh{0.70}, \Thresh{0.75} and MLP.  While MLP gives the best performance, the threshold models still provide very strong performance and present a clear and intuitive tradeoff between precision and recall.

\begin{table}[htbp]
\centering
\begin{tabular}{c|ccc|ccc|l}
\toprule
& \multicolumn{3}{|c|}{Synthetic Data} & \multicolumn{3}{|c|}{Manual Match Data} & \\
Model & Prec. & Recall & F1 & Prec. & Recall & F1 & Model Parameters \\
\midrule
\multicolumn{8}{c}{Bloom Filter: 2-ary Q-gram, 64 bit vectors}\\
\midrule
Tree & 0.8855 & 0.796 & 0.8384 & 0.9995 & 0.8443 & 0.9153 &  5 max leaf nodes \\
LR & 0.8123 & 0.8246 & 0.8184 & 1.0 & 0.8658 & 0.928 & w=1:100,C=$10^{-5}$,liblinear,l1 \\
MLP & 0.9446 & 0.9118 & 0.9279 & 0.998 & 0.9279 & 0.9617 & adam,$\alpha$=$10^{-6}$,layers=(15,) \\
\Thresh{0.70} & 0.9273 & 0.9009 & 0.9139 & 0.9986 & 0.9498 & 0.9736 & $\beta$=0.70 \\
\Thresh{0.75} & 0.9867 & 0.841 & 0.908 & 1.0 & 0.8945 & 0.9443 & $\beta$=0.75 \\
\midrule
\multicolumn{8}{c}{Bloom Filter: 2-ary Q-gram, 32 bit vectors}\\
\midrule
Tree & 0.8732 & 0.6016 & 0.7124 & 0.9993 & 0.669 & 0.8014 & 5 max leaf nodes \\
LR & 0.2009 & 0.9317 & 0.3305 & 0.7748 & 0.9778 & 0.8645 & w=1:100,C=$10^{-5}$,liblinear,l1 \\
MLP & 0.9593 & 0.8603 & 0.9071 & 0.9973 & 0.9059 & 0.9494 & adam,$\alpha$=$10^{-6}$,layers=(15,) \\
\Thresh{0.75} & 0.9087 & 0.8806 & 0.8944 & 0.9995 & 0.9393 & 0.9685 & $\beta$=0.75 \\
\Thresh{0.80} & 0.9867 & 0.8138 & 0.8919 & 1.0 & 0.8598 & 0.9246 & $\beta$=0.80 \\
\bottomrule
\end{tabular}
\caption{Pairwise linkage test performance.}
\label{tb.PairPerf}
\end{table}

Table~\ref{tb.ClstrPerf} presents the performance of the CENTER cluster (CC) and CENTER-MERGE cluster (CMC) algorithms using the precision and recall metrics defined in Section~\ref{ssec:Cluster}.  There is a clear precision/recall tradeoff between the CC and CMC algorithms.  This is consistent with \cite{hassanzadeh2009a} where the authors describe the CMC algorithm as being based on CC with the additional mechanism to merge clusters that share a common node.  This serves to create larger clusters which improves recall at the expense of precision.  When comparing the \Thresh{0.70} and \Thresh{0.75} results in Table~\ref{tb.ClstrPerf}, the higher threshold achieves higher precision with lower recall as expected.  While the MLP algorithm offers acceptable performance, \Thresh{0.75} does better in spite of MLP being the best in Table~\ref{tb.PairPerf}.  An important input to both CC and CMC are the confidence values associated with each link.  It is possible that the output of the {\tt perf\_proba()} function influenced the creation of the MLP clusters relative to the more direct $D_{\rm All}$ Dice coefficient confidence calculation for the threshold methods.

\begin{table}[htbp]
\centering
\begin{tabular}{r|ccc}
\toprule
Link Alg., Cluster Alg. & Precision & Recall & F1 \\
\midrule
\Thresh{0.75},CC & 0.948 & 0.936 & 0.942 \\
\Thresh{0.75},CMC & 0.924 & 0.953 & 0.938 \\
\Thresh{0.70},CC & 0.864 & 0.930 & 0.896 \\
\Thresh{0.70},CMC & 0.686 & 0.970 & 0.804 \\
MLP,CC & 0.918 & 0.902 & 0.910 \\
MLP,CMC & 0.880 & 0.922 & 0.901 \\
\bottomrule
\end{tabular}
\caption{Cluster algorithm performance on manually matched clusters.}
\label{tb.ClstrPerf}
\end{table}

To investigate this and to offer a perspective on the behaviour of the linkage/clustering algorithms on the entire $K$=235,230 data set, Table~\ref{tb.ClstrScores} shows the mean, median and minimum pairwise linkage confidence scores, described in Section~\ref{ssec:PairMLModels}, for all the clusters identified by the CC and CMC algorithms.  While the median confidence value is 1.0 in all cases, the MLP algorithm has the highest average confidence in spite of the linkage probability having a minimum value of 0.5 and the threshold confidences being lower bounded by their threshold values.  It is possible that the MLP model is too confident for some of its links which led to its degradation in performance as shown in Table~\ref{tb.ClstrPerf}.  Cluster sizes created by linking the full data set are shown in Tables~\ref{tb.ClstrSzCC} and \ref{tb.ClstrSzCMC} for the CC and CMC algorithms, respectively.  These results can be compared with the manually linked cluster sizes in Table~\ref{tb.ClstrSz}.  It can be noted that cluster sizes for the full data set trend smaller with a higher percentage of single profile clusters and a lower percentage of clusters with more than 5 profiles.  This is a result of the recall limitations of the linkage methods used.

\begin{table}[htbp]
\centering
\begin{tabular}{rccc}
\toprule
Link Alg., Cluster Alg. & Mean & Median & Min \\
\midrule
\Thresh{0.75},CC & 0.955 & 1.000 & 0.750 \\
\Thresh{0.75},CMC & 0.955 & 1.000 & 0.750 \\
\Thresh{0.70},CC & 0.927 & 1.000 & 0.700 \\
\Thresh{0.70},CMC & 0.934 & 1.000 & 0.700 \\
MLP,CC & 0.971 & 1.000 & 0.500 \\
MLP,CMC & 0.974 & 1.000 & 0.500 \\
\bottomrule
\end{tabular}
\caption{Cluster confidence.}
\label{tb.ClstrScores}
\end{table}

\begin{table}[htbp]
\centering
\begin{tabular}{c|c|c|c}
\toprule
Cluster & & & \\
 Size & \Thresh{0.70} & \Thresh{0.75} & MLP \\
\midrule
1 & 26,306 (29.29\%) & 31,161 (34.70\%) & 30,538 (34.00\%)\\
2 & 26,837 (29.88\%) & 33,544 (37.35\%) & 32,492 (36.18\%)\\
3 & 11,140 (12.40\%) & 12,753 (14.20\%) & 12,659 (14.10\%)\\
4 & 7,727 (8.60\%) & 8,241 (9.18\%) & 8,305 (9.25\%)\\
5 & 4,830 (5.38\%) & 4,861 (5.41\%) & 4,868 (5.42\%)\\
$>$5 & 8,974 (9.99\%) & 6,355 (7.08\%) & 6,627 (7.38\%)\\
\bottomrule
\end{tabular}
\caption{CENTER algorithm cluster sizes.}
\label{tb.ClstrSzCC}
\end{table}

\begin{table}[htbp]
\centering
\begin{tabular}{c|c|c|c}
\toprule
Cluster & & & \\
 Size & \Thresh{0.70} & \Thresh{0.75} & MLP \\
\midrule
1 & 22,009 (24.51\%) & 29,434 (32.77\%) & 28,333 (31.55\%)\\
2 & 23,395 (26.05\%) & 32,465 (36.15\%) & 31,002 (34.52\%)\\
3 & 9,268 (10.32\%) & 12,049 (13.42\%) & 11,727 (13.06\%)\\
4 & 5,567 (6.20\%) & 7,443 (8.29\%) & 7,204 (8.02\%)\\
5 & 3,392 (3.78\%) & 4,445 (4.95\%) & 4,284 (4.77\%)\\
$>$5 & 6,816 (7.59\%) & 7,112 (7.92\%) & 7,261 (8.08\%)\\
\bottomrule
\end{tabular}
\caption{CENTER-MERGE algorithm cluster sizes.}
\label{tb.ClstrSzCMC}
\end{table}

\subsection{System of Care Utilization}
\label{ssec:SysUse}

Tables~\ref{tb.Stays}, \ref{tb.Tenure} and \ref{tb.Visits} show total shelter stays, shelter tenure and shelters visited per person, respectively.  Each table shows mean and median results for two cohorts: all of the 85,229 people in the shelter data (All) and the top 5th percentile of those people in terms of the table metric (Top 5\%).  It is important to examine this top percentile since it is often a small minority of people within an HHSC that make the heaviest (ie. chronic) use of the system \cite{kuhn1998,culhane2007,aubry2013,kneebone2015}.  Results are shown for the un-merged data set and the six pairwise linkage/clustering algorithm combinations.

\begin{table}[htbp]
\centering
\begin{tabular}{cc|c|c|c|c|c|c|c}
\toprule
 & & Un-  & \multicolumn{2}{|c|}{\Thresh{0.70}} & \multicolumn{2}{|c|}{\Thresh{0.75}} & \multicolumn{2}{|c}{MLP} \\
 Cohort & &  Merged  & CC & CCM & CC & CCM & CC & CCM \\
\midrule
All & Mean & 97.8 & 125.3 & 162.8 & 117.2 & 122.6 & 117.6 & 125.7\\
& Median & 11.0 & 16.0 & 16.0 & 14.0 & 15.0 & 14.0 & 15.0\\ \hline
Top & Mean & 949.7 & 1142.2 & 1922.5 & 1093.8 & 1146.8 & 1095.7 & 1195.6\\
5\% & Median & 846.5 & 1023.0 & 1042.0 & 979.0 & 1015.5 & 978.0 & 1026.0\\
\bottomrule
\end{tabular}
\caption{Total shelter stays per person.}
\label{tb.Stays}
\end{table}

\begin{table}[htbp]
\centering
\begin{tabular}{cc|c|c|c|c|c|c|c}
\toprule
 & & Un-  & \multicolumn{2}{|c|}{\Thresh{0.70}} & \multicolumn{2}{|c|}{\Thresh{0.75}} & \multicolumn{2}{|c}{MLP} \\
 Cohort & &  Merged  & CC & CCM & CC & CCM & CC & CCM \\
\midrule
All & Mean & 358.1 & 439.6 & 410.8 & 399.2 & 407.2 & 406.0 & 414.2\\
& Median & 42.0 & 72.0 & 53.0 & 54.5 & 55.0 & 56.0 & 57.0\\ \hline
Top & Mean & 1967.7 & 2061.9 & 2056.1 & 2021.3 & 2040.8 & 2027.2 & 2051.1\\
5\% & Median & 1938.0 & 2028.5 & 2027.0 & 1990.0 & 2010.0 & 1994.5 & 2020.0\\
\bottomrule
\end{tabular}
\caption{Shelter tenure (days).}
\label{tb.Tenure}
\end{table}

\begin{table}[htbp]
\centering
\begin{tabular}{cc|c|c|c|c|c|c|c}
\toprule
 & & Un-  & \multicolumn{2}{|c|}{\Thresh{0.70}} & \multicolumn{2}{|c|}{\Thresh{0.75}} & \multicolumn{2}{|c}{MLP} \\
 Cohort & &  Merged  & CC & CCM & CC & CCM & CC & CCM \\
\midrule
All & Mean & 1.2 & 1.4 & 1.4 & 1.3 & 1.4 & 1.3 & 1.4\\
& Median & 1.0 & 1.0 & 1.0 & 1.0 & 1.0 & 1.0 & 1.0\\ \hline
Top & Mean & 2.4 & 3.3 & 3.4 & 3.2 & 3.3 & 3.2 & 3.3\\
5\% & Median & 2.0 & 3.0 & 3.0 & 3.0 & 3.0 & 3.0 & 3.0\\
\bottomrule
\end{tabular}
\caption{Shelters visited per person.}
\label{tb.Visits}
\end{table}

\section{Discussion}
\label{sec:Discussion}

This paper demonstrates that privacy preserving linkage techniques can be successfully applied in the housing and homelessness system.  The practical implications of this are significant.  As discussed in Section~\ref{sec:Intro}, data fragmentation and the lack of a unique identifier for system users are fundamental properties of most HHSCs.  However, understanding the flow of a population through that system and how different agencies are or are not working well together is an essential first step towards improving care.  The privacy preserving linkage methods demonstrated in this paper are an effective way of creating a system-wide HHSC data set that is considerably cheaper and more practical than migrating all the agencies within an HHSC to a common data system and providing all system users with a unique identifier.  The linked data remains anonymous with our approach which does not allow agencies within in HHSC to look up a specific person's linked records.  However, this possible limitation is also an advantage since it frees agencies from implementing the rigorous consent process that would be required for plain text identifying information to be shared between agencies.

The results in Section~\ref{sec:Results} also demonstrate the importance of evaluating a record linkage scheme with a variety of metrics at each stage of the entity resolution process.  While MLP offered the best performance at the pairwise linkage level in Section~\ref{ssec:AlgPerf}, the simpler threshold detection method proved to be the best after the second clustering step.  Including performance metrics specific to the application domain, in this case housing and homelessness, is also important.  In some ways, the performance results in Sections~\ref{ssec:AlgPerf} and \ref{ssec:SysUse} are consistent.  For example, the techniques in Table~\ref{tb.ClstrPerf} with lower F1 scores tend to maintain a reasonable recall level with more significant drops in precision.  This suggests the methods are still creating large clusters with an increasing number of false positives.  These larger clusters would be consistent with the larger total shelter stay and tenure numbers in Tables~\ref{tb.Stays} and \ref{tb.Tenure}, respectively, when comparing \Thresh{0.70} (low F1) with \Thresh{0.75} (high F1).  

A closer comparison of Sections~\ref{ssec:AlgPerf} and \ref{ssec:SysUse} reveals that both machine learning and domain specific metrics need to be considered when selecting a final linkage algorithm.  While the \Thresh{0.75},CC combination yields the best F1 score in Table~\ref{tb.ClstrPerf}, the possible downside of selecting one of the other alternatives from that table doesn't seem significant when looking at F1 score alone.  For example, the penalty of going with a sub-optimal threshold, for example \Thresh{0.70},CC, results in a percentage drop in F1 of only 4.9\%.  A machine learning engineer could take these results to mean that time could be saved by hyperparameter tuning the threshold in coarser steps in the future.  However, the percentage difference between \Thresh{0.75},CC and \Thresh{0.70},CC when examining HHSC system use tenure is doubled to 10\% or approximately 40 days.  This increased sensitivity in domain specific metrics is important to note and discuss with the housing and homelessness program designers who would be the ultimate users of the results in Section~\ref{ssec:SysUse}.

Finally, regardless of the linkage algorithm chosen, the results in Section~\ref{ssec:SysUse} indicate that some form of data linkage is important when understanding HHSC system use.  If \Thresh{0.75},CC is chosen as the linkage method based on its superior F1 score, comparing the \Thresh{0.75},CC results to the un-merged data in Tables~\ref{tb.Stays}, \ref{tb.Tenure} and \ref{tb.Visits} shows significant change.  For example, the median results for shelter stays increases by over 100 days (15.7\%) for the top 5th percentile of shelter users when using linked data instead of unlinked data.

\section{Conclusion}

This paper demonstrates that privacy preserving record linkage is both practical for the housing and homelessness sector and necessary for understanding how a person interacts with the overall housing and homelessness system of care.  The picture of how people use an HHSC changes when working with a fully linked data set.  The specific algorithms appropriate for performing this linkage are also surprisingly straightforward.  The best overall linkage performance is be achieved by applying a simple threshold to Dice coefficient values calculated from Bloom filter scrambled identifying information followed by a computationally efficient clustering scheme to link profiles to latent individuals.  Observing the effect of linkage algorithms on system utilization metrics specific to the housing and homelessness sector reveals that linkage has a measurable effect on a population's picture of HHSC use. These same domain specific metrics also play an important role in evaluating the best algorithm combination to perform the linkage.

\section{Conclusion}
The author would like to acknowledge the support of Making the Shift, the Calgary Homeless Foundation and the Government of Alberta.  This study is based in part on data provided by Alberta Seniors, Community and Social Services.  The interpretation and conclusions contained herein are those of the researchers and do not necessarily represent the views of the Government of Alberta.  Neither the Government of Alberta nor Alberta Seniors, Community and Social Services express any opinion related to this study.

\bibliographystyle{ACM-Reference-Format}
\bibliography{ggmessier-lib}

\end{document}